\documentstyle{elsart}
\def\su#1{\mbox{$\mathrm{SU}(#1)$}}
\def\so#1{\mbox{$\mathrm{SO}(#1)$}}

\def\msu#1{\mbox{\mathrm{SU}(#1)}}
\def\at{\mbox{$A^\mathrm{\tau}$}} 
\def\ae{\mbox{$A^\mathrm{e}$}}

\def\afbt{\mbox{$A^\mathrm{\tau}_\mathrm{FB}$}}

\def\alr{\mbox{$A_\mathrm{LR}$}}
\def\dt{\mbox{$\delta$}}
\def\mt{\mathrm}
\def\i{\mbox{\mt{i}}}
\def\j{\mbox{\mt{j}}}
\begin{document}
\rightline{OUTP 9655P}
\begin{frontmatter}
\title{\bf The $\tau$ forward-backward asymmetry within Grand Unified 
Theories} 
\author{Francesco Caravaglios \thanksref{EC}}
\address{Theoretical Physics, University of Oxford, 1 Keble Road,
Oxford OX1 3NP, UK}
\thanks[EC] {European Community Fellow, EC contract
no. ERBCHBGCT940685.\\
Address from November 96: TH-Division,  CERN, Geneva.}

\begin{abstract}
We will study some possible  non universal corrections that could
affect single measurements of the Weinberg angle $\sin^2
\theta_\mt{W}$ (leaving the full average essentially unchanged).
We will concentrate on the class of the models that introduces a new
light gauge boson and an extended Higgs sector, but we restrict
the Higgs charges choice in order to {\it automatically} achieve zero mixing
at  the tree level between the $Z_0$ and the $Z^\prime$. 
This will more naturally keep small the universal effects with a
relatively light $Z^\prime$, but it will
imply only vector-like coupling between this boson and the standard
fermions. A \so{10} gauge boson which distinguishes between families
can have such properties and could explain a deviation in the \afbt
without affecting the $\tau$ polarization measurements at LEP. 
\end{abstract}
\end{frontmatter}
\section{Introduction}
The analysis of the electroweak physics  at the $Z_0$ peak has confirmed
 the agreement between the Standard Model (SM) and the experiments beyond
 a trivial level of accuracy [1]. One loop pure electroweak 
corrections,
involving the  top quark and the Higgs boson sector  have  been tested.
The top mass measured by CDF/D0 strengthen this success.
The experimental result for the Weinberg angle is obtained taking the 
average of all the
available precise measurements.
The global average between very different type of measurements (both hadronic
and leptonic) is justified by the fact that any new physics [2]
 contributing
to  the vacuum polarization of the weak bosons 
(the most likely place where to look for 
new effects) will affect in a universal way all the precise 
measurements [1].
Nevertheless this does not exclude the possibility 
of one single measurement being affected while leaving the whole average
essentially unchanged.

In the following we discuss how an additional gauge boson [3]
 could change
only one (or some)  of the forward-backward asymmetries and we try to
embed it in a Grand Unified scenario.

\section{Electroweak precise measurements: the $\tau$ forward-backward
 asymmetry}
The forward-backward asymmetry of the $\tau$ lepton deviates from the
Standard Model of about $2~\sigma$. On the contrary the $\tau$
polarization measurements agree with the SM; an anomalous coupling
between the $Z_0$ and the $\tau$ cannot explain their discrepancy.
But if we add to the SM  scattering amplitude
\begin{equation} 
\label{ampsm}
e^++e^- \rightarrow Z_0\rightarrow \tau^+ + \tau^-
\end{equation}
a new contribution 
\begin{equation} 
e^++e^- \rightarrow X\rightarrow \tau^+ + \tau^-
\end{equation}
  which is imaginary and thus it interferes with the SM amplitude 
induced by  the $Z_0$ production,  we could affect the $A_{FB}^\tau$ 
keeping negligible effects  to all the other precise measurements. 
For instance, let us assume that this $X$ is  a new $Z^\prime$.
In a physically interesting class of models  described in 
 the next section the mixing between  the $Z_0$ and the $Z^\prime$ can be
very small, so  we will consider it to be negligible\footnote{The
effects of the mixing of a $Z^\prime$ with a $Z_0$ are extensively
discussed in the literature.}. In these models the coupling of
this new boson with the standard fermions can only be vector-like (see
next section).
We fix the interaction between the $\tau$  and the boson to be 
\begin{equation}
g_\tau Z^\prime_\mu \bar\tau \gamma^\mu \tau.
\end{equation}  
This will define the coupling constant $g_\tau$. Analogously we define
$g_e,g_\mu,\dots$ the couplings with the electron, the muon etc.
This $Z^\prime$ will add to the above scattering amplitude 
a term  ($p^2=M_Z^2$)
\begin{equation}
g_e g_\tau~ \bar e \gamma_\mu e \bar \tau\gamma^\mu \tau  {1\over
M^2_Z-M_{Z^\prime}^2+ \mt{i} M_Z \Gamma_{Z^\prime}}
\end{equation}
whose imaginary part is ($\Gamma_{Z^\prime}<<M_Z -M_{Z^\prime}$)
\begin{equation} 
\label{amp}
g_e g_\tau ~\bar e \gamma_\mu e \bar \tau\gamma^\mu \tau\times  {\mt{i} M_Z
\Gamma_{Z^\prime}
\over (M_Z^2-M_{Z^\prime}^2)^2}=\mt{i}
g_e g_\tau ~\bar e \gamma_\mu e \bar \tau\gamma^\mu \tau  \times
{C\over M_Z^2}
\end{equation}
where for simplicity we have called $C$ the factor that includes the
mass and the width of the new boson.
This imaginary part will affect the precise  measurements of the
$\tau$ lepton.
Computing the interference between  the above matrix element
(\ref{amp})
and  the   SM amplitude (\ref{ampsm}) we obtain that 
the correction to the total cross section $\tau^+ ~\tau^-$
   is proportional  to the 
product of the vector  couplings $g_V^e$ and  $g_V^\tau$ 
 of the $Z_0$ with the electron and the $\tau$.
These parameters are small and the effect  is suppressed and negligible. 
Analogously the polarization measurement of the $\tau$ gets
corrections which are proportional to  $g_V^e$ or  $g_V^\tau$, and
thus are  very small.
On the contrary the forward backward asymmetry of the $\tau$ gets a
correction which is proportional to $g_A^e$ and $g_A^\tau$ and
therefore is numerically sizable.
This argument is true only because  we have  assumed our new  gauge boson  to
be vector-like (hereafter we will call   vector-like a gauge boson with purely
vector-like interaction with the standard fermions, see also the next
section).
In table (\ref{tab}) we show how the measurements of the Weinberg angle
involving the $\tau$ lepton are affected by   the interference between
the   amplitude (\ref{amp}) and  the SM matrix element (\ref{ampsm}).
\begin{table}
\center{\caption{\label{tab}Effects to the electroweak asymmetries.}
\begin{tabular}{|c|c|c|c|}
\hline
\hline
\dt\afbt  &\dt \alr$^\mt{slac}$ & 
\dt\ae$_\mt{pol}$  & \dt\at$_\mt{pol}$ \\
\hline
 1.25  $g_\mt{e}$  $g_\tau$ $C$  & 0.
 & 0.12  $g_\mt{e}$ $g_\tau$ $C$ &0.12 $g_\mt{e}$ $g_\tau$  $C$\\
\hline
\hline
\end{tabular}}
\end{table}

We have seen that an imaginary amplitude such as in (\ref{amp}) can
only affect  the $A_{FB}^\tau$;  now we investigate in some details 
the specific case of a Grand Unified boson as described in the last
section. This boson (see last section) is coupled at tree level only 
with the third generation
of fermions.
If the coupling with the $\tau$ is $g_\tau=g$ then from the  table
(\ref{coupl})
we have $g_b=g_t=-1/3 g $ and $g_{\nu_\tau}=g$.
Even if the tree level coupling of this boson with the electron is
zero, at one loop a small coupling can arise.
For instance the two-point Green function $Z^\prime-\gamma$ 
is generated\footnote{A
small electron coupling could also arise through  a two-point function 
$Z^\prime-Z^{\prime\prime}$ where $Z^{\prime\prime}$ is a vector-like
boson much heavier than the $Z^\prime$ and coupled to the electron.} by  the
one loop fermionic corrections.
The imaginary part of this function gives an amplitude similar to the 
(\ref{amp})
\begin{equation}
\label{ampgam}
  e_\gamma g~ \bar e \gamma_\mu e \bar \tau\gamma^\mu \tau\times  {\mt{i}
A_{\gamma Z^\prime}
\over M_Z^2 (M_Z^2-M_{Z^\prime}^2) }
\end{equation}
where $e_\gamma$ is the electromagnetic coupling, $1/M_Z^2$ and 
$1/(M_Z^2-M_{Z^\prime}^2)$ come from the photon and the $Z^\prime$
propagators;
  to compute the constant $A_{\gamma Z^\prime}$ we have to
evaluate the imaginary part of the fermionic loop which mixes the photon
and the $Z^\prime$ propagators. 
The only fermions contributing to the imaginary part in this loop 
 are the bottom and the $\tau$ since the top quark is above the
threshold. The
couplings have been given previously.
The only measurement affected would be the $A_{FB}^\tau$;  the
correction can be computed  from table (\ref{tab}) replacing   the factor 
$g_e C$ in  (\ref{amp}) with the proper one in (\ref{ampgam});
 calculating the  loop $A_{\gamma Z^\prime}$ we obtain 
\begin{equation}
\delta A_\mt{FB}^\tau= 0.0002\times 
{g^2\over e^2}\times {M_Z^2 \over M_Z^2-M^2_{Z^\prime}}
\end{equation} 
This could be sizable if the $Z^\prime$ is not too far from the $Z_0$
(the 
limits for such a boson essentially coupled only to the third family
and decaying mainly  in $\tau$  or invisible/dark
matter  are weak).
All the four LEP experiments give an excess in the \afbt 
 (see  table (\ref{exp})), but the average is only $2\sigma$ away from
the SM. 
Before concluding this section, it is worthwhile to note again  that, 
differently from the $\gamma-Z^\prime$, a  
$Z_0-Z^\prime$ mixing  cannot  explain the discrepancy between
 $A_{FB}^\tau$ and $A^\tau$: the effect would be the same in both 
measurements, since it could be reabsorbed into a redefinition of the
$\tau$ coupling with the $Z_0$;    to compute such a correction 
 we should evaluate the
real part of the fermionic loop, which depends on   the spectrum
 (and   its several free parameters) of 
the theory  above the $Z_0$ mass   
(remembering that the $Z_0$ is purely
imaginary and the photon is real at the $Z_0$ peak, the phase of the
fermionic loop must be chosen accordingly).

\begin{table}
\center{\caption{\label{exp} The $\tau$ forward-backward asymmetry compared with
the SM value.}
\begin{tabular}{|c|c|c|c|}
\hline
\hline
EXP.& DATA & LEP  AVERAGE & STANDARD MODEL   \\
\hline
ALEPH &$ 0.0196\pm0.0028$ &$ 0.0201\pm0.0018$ & $ 0.0160 $\\

DELPHI &$ 0.0223\pm0.0039$ &$ 0.0201\pm0.0018$ & $ 0.0160 $\\

L3 &$ 0.0233\pm0.0049$ &$ 0.0201\pm0.0018$ & $ 0.0160 $ \\

OPAL &$0.0178\pm0.0034$  &$ 0.0201\pm0.0018$ & $ 0.0160 $\\
\hline
\hline
\end{tabular}}
\end{table}
\section{Adding a new gauge boson}
Within the SM the $Z_0$ and $W$ mass are related at tree level by a
well known relation  
\begin{equation}
\label{rel}
M_Z^2=M_W^2 { g_1^2+g_2^2 \over g_2^2}.
\end{equation}
This give a prediction of the $W$ mass once the $Z_0$ mass and the
weak coupling are known, if we use a minimal representation of 
Higgs sector: {\it i.e.} one (or more) electroweak doublet. 

If we add an extra massive gauge boson to our model the Higgs sector must be
enlarged, since the longitudinal polarization of the new massive boson
must come from the  would-be goldstone boson of an additional scalar
representation.
A new Higgs must acquire a $vev$, which must carry a non zero
charge with respect the new gauge boson.
We briefly discuss two possible scenarios:\\
\subsection{non minimal low energy  Higgs sector} 
 at least one $vev$  carries 
charges both of \su2$\times \mt{U}(1)$ and $\mt{U}^\prime(1)$.
As a  consequence this $vev$ will introduce a mixing between the $Z_0$
and the $ Z^\prime$ proportional  to 
\begin{equation}
 v^2 Z_0 Z^\prime.
\end{equation}   
The physical mixing between the $  Z_0$ and the $ Z^\prime$ can be small
if we add another $vev$ 
   $v^\prime>>v,v_{SM}$ neutral with respect the SM
gauge generators but with a   $ Z^\prime$ charge:
the new boson becomes very heavy  and the mixing negligible.
In this case the relation (\ref{rel}) is a consequence of a particular
choice of the Higgs potential, where two Higgs acquire $vev$'s of  different   
order of magnitude.
Since these two fields share the $\mt{U}^\prime(1)$ interaction, one
has probably  to introduce a mechanism to explain this hierarchy.
\subsection{ minimal low energy  Higgs sector: a vector-like gauge
boson }
Another possible scenario arises if we just add  a minimal extension
of the Higgs sector of the low energy effective lagrangian.
In such a case we add only one Higgs scalar which is   a SM singlet
and  coupled only to the $Z^\prime$, while the usual standard model
Higgs doublet is  not coupled to the $Z^\prime$.  
Only this charge choice of the scalars  guarantees that no
mixing arises at tree level between the $Z_0$ and the $Z^\prime$.
As for the SM the  tree level relation (\ref{rel}) is simply a
consequence   of the Higgs charge choice, regardless the shape of the
Higgs potential and its arbitrary parameters.
 Differently from the first case, now the new gauge
boson can be more easely much closer to the weak scale, since no assumption on the
Higgs potential and no hierarchy is
required. 

The models with a minimal extension of the Higgs sector of the  
low energy effective lagrangian have an important property in addition 
to the tree level equation (\ref{rel}).   
The additional gauge boson can  only have a vector-like interaction with the
 dirac fermions of the standard model:
each standard fermion (having the standard charge assignments with
respect \su2$\times\mt{U}(1)$) can get mass  only through the 
\su2 Higgs doublet.
 Remembering that this does not  carry $Z^\prime$ charge (see above),
a gauge  invariant Yukawa interaction between the dirac fermion
 and the Higgs doublet is allowed only if the left and right handed
component of the dirac fermion transform with the same charge with
respect  $\mt{U}^\prime (1)$ gauge transformation.
In other words the coupling between this   $\mt{U}^\prime (1)$ and the
standard fermions is vector-like.

In this work we have concentrated on the latter  class of minimal
models 
where the (\ref{rel}) is a direct consequence of the Higgs
representation choice, regardless the Higgs potential shape of the
low energy effective lagrangian; in the last section we will 
comment on   the consequences of the embedding such a gauge boson in a 
Grand Unified model.

\section{Grand Unified Theories}
All the discovered gauge bosons have couplings with the fermions which
fit in a very nice and simple way with a \su5 Grand Unified minimal
scenario [4].
Therefore a new $Z^\prime$ that can be easely  accommodated in  the above
picture is certainly welcome.  
The experimental constraints for (light) Grand Unified extra $\mt{U}(1)$ are
in general very compelling. If they have a tree level  coupling 
 with all the existent fermions then strong constraints come from
CDF/UA2: its direct production  would certainly have been  detected
in the leptonic decay modes. 
But if its leptonic coupling is suppressed and/or the
light quark sector is decoupled, this gauge boson could
be very light, even close to the weak scale, probably escaping all  the direct 
searches. 

In the following we will discuss how such a gauge boson could be
accommodated in a Grand Unified scenario.

\subsection{A GUT gauge boson}
If both the left handed  up quark
 and the charge conjugated of the
right handed up quark   belong to the same
irreducible representation  (of a Grand unified group) 
then a vector-like gauge boson 
has opposite couplings with the above states of the same
representation: its generator  cannot commute  for instance 
with  a \su5 embedding the SM. 
Equally,
if we require leptophobia  
the gauge boson cannot commute with this group since  
the  electron components (left and  right handed)
 and the  quarks   are contained in  the  same irreducible
representations of the unified group.
Conversely the generator of this gauge boson must commute with 
 $\mt{SU}(3)\times\mt{SU}(2) \times \mt{U}(1)$: it cannot carry color
(one of the reasons is because  we want it to have a small but non zero
coupling with the electron, for colored gauge bosons see after),
 we do not want it to carry    $\mt{SU}(2) \times \mt{U}(1)$ charges,
it must not have neither electromagnetic charge nor weak isospin
charge, otherwise it would be or a charged current or at least a
member of 
 a $\mt{SU}(2)$ multiplet with a charged boson (the splitting within
this  $\mt{SU}(2)$ multiplet would be bounded by the electroweak
precise measurements, {\it e.g.} $\varepsilon_1$).
Since the generator of the new boson commutes with \su2, the lefthanded
components of the top and bottom quarks must have the same charge  $g_q$.
From the vector-like nature of the new boson we obtain that also their
righthanded components  have charge $g_q$.
For the same reasons all the leptons have the same charge $g_l$.
The anomaly cancellation (more precisely 
the square of the hypercharge times the new generator)
  will force us to fix the charges as in table (\ref{coupl}) (a
righthanded neutrino is also required).
This charge choice corresponds to a linear combination of the two
$U(1)\times U(1)^\prime$ in \so{10}.
It is manifest that this \so{10} cannot be the standard one containing
the SM gauge bosons, otherwise this new boson 
 would be coupled to the electron, and
certainly (if not too heavy)  it would have been  seen at CDF/D0.
Therefore it has to  belong to  an additional \so{10}$_{NEW}$: the
standard \so{10} is coupled to all  the three fermion families while 
the additional  \so{10}$_{NEW}$ 
is not coupled with the two lightest families.
  
For example,one could be   attracted by the idea that some  family generation
of gauge bosons could exist as well as the generations of fermion families.

If the existence of fermion family copies is  not understood 
in the context of a simple  Grand Unified model one could suspect that the
answer  to this problem is beyond the unification scale and the
mechanism (generating the families) could not distinguish between
fermion and 
gauge boson but
could simply give some copies of the same gauge lagrangian 
\begin{equation}
{ L=L_1+L_2+L_3+...}~~~~~~~~~~~1,2,3,\dots=~\mt{indices~ of~ generations}
\end{equation} 
The gauge bosons would have a family index exactly as the fermions have.
To reduce the number of non-abelian groups to
the SM \su3$\times$\su2, we  need representations obtained from the
tensorial  products: for instance,  
any Higgs field $\phi^{\alpha,\beta} $ belonging to the $ (8_\mt{i},8_\mt{j})$ of 
 $\msu3_{\i} \times \msu3_{\j} $ could acquire a $vev$ 
 $\phi^{\alpha,\beta}=v\delta_\mt{kronecker}^{\alpha,\beta}$.
This will break $\msu3_{\i} \times \msu3_{\j} \Rightarrow 
\su{3}_{\i+\j}$.
The same breaking can occur for the \su2.

After all the non abelian groups are broken into the SM ones,
the low energy model  would have  the SM gauge bosons
 but in addition one could 
have an  abelian gauge boson (remained unbroken at the
unification scale) which is coupled only to the third  (and to an  extra 
heavy vectorial family, see below).
We would have   three   generation
of fermions at the weak scale plus 
another heavier vectorial 
family in order to satisfy  the orthogonality condition 
$\mt{Tr} T_\mt{SM} T^\prime=0$,   with $T_\mt{SM}$  any SM gauge boson
generator and $T^\prime$ the boson in table (\ref{coupl}).
Other light {\it dark} generations not
coupled with the standard boson but coupled with the $Z^\prime$ 
 obviously cannot be ruled out.
The overall  normalization of the effective low energy  coupling in
table  (\ref{coupl}) 
is a   free parameter, 
and  it also depends on
the total number of families coupled to the boson and the  details of
the symmetry breaking at the GUT scale.
\begin{table}
\center{\caption{\label{coupl}  Couplings of the vector-like
 gauge boson to the 16 of \so{10}$_{NEW}$.}
\begin{tabular}{|c|c|c|c|}
\hline
\hline
$\mt{t}$ & $\mt{b}$ & $\tau$ & $\nu_{\tau}$ \\
\hline
 $-{1\over 3}$&$-{1\over 3}$&$ {1}$ &$ {1}$  \\
\hline
\hline
\end{tabular}}
\end{table}
Finally,  we point out that its vector-like nature will easely allow an
electroweak breaking which automatically preserves  the tree level relation
between the $Z_0$ and the $W^\pm$ masses. 
It is clear from table (\ref{coupl}) that it is possible to give mass
to the third generation with a Higgs neutral under the $Z^\prime$,
in a GUT scenario it could be the doublet in the 10 of \so{10}, whose
charge with respect the generator of table (\ref{coupl}) is zero. 
On the contrary the $Z^\prime$ can get mass from the SM singlet in the
16 of \so{10}$_{NEW}$. No Higgs  that is coupled both to the $Z_0$ and the
$Z^\prime$ acquire a $vev$ and the  mixing between the two gauge bosons
is zero at the tree level.
At one loop a small mixing can arise between the two gauge bosons.
Similarly, the radiative corrections could introduce a higher order
operator 
\begin{equation} 
\label{oper}
H_2 H_0 b_L s_R 
\end{equation}
that can explain a small mixing between the third generation and the
second one.
Since  both the strange quark $s_R$ and  the Higgs doublet $H_2$ are
neutral under $Z^\prime$, the  additional Higgs $vev$ $H_0$ is necessary
 to build a gauge invariant mixing  in  the (\ref{oper}).
It carries only a $Z^\prime$ charge and 
 this will  avoid  a tree level $Z_0-Z^\prime$ mixing.    
\section{Conclusions}
After the success of the precision tests of the {\it universal} corrections to the observables at the $Z_0$ peak, we have tried to look for possible signals of new physics in single measurements of the $\sin^2 \theta_\mt{w}$.
Given that the electroweak relation 
\begin{equation}
M_\mt{Z}^2={g_1^2+g_2^2\over g_2^2} M^2_\mt{W}
\end{equation}
is remarkably true, we have studied the special class of models in which,
as for the SM, the above relation is a direct consequence of the 
Higgs representation choice, with  
no assumption on the free parameters of the Higgs potential.
These models demand that the new $Z^\prime$ has a vectorial
interaction 
with the standard fermions.
Thus deviations in  the  forward-backward  asymmetries could signal
the existence of such bosons.

We have also discussed   this $Z^\prime$ in the context of 
Grand Unified Theories.
If we demand that the $Z^\prime$ has a suppressed coupling with the
electron 
and its  tree level mixing with the $Z_0$ is automatically and
naturally zero
 (see above), 
we are practically forced to introduce a  new generation of gauge bosons.

This could belong to the adjoint of \so{10}.
This group contains a  vector-like gauge boson, which 
  is a linear combination of 
the   $\mt{U}(1)\times\mt{U}^\prime(1)$ commutating with 
\su3$\times$\su2.
If this gauge boson is coupled with only the third generation of
 fermions it could escape all the direct searches even if it is light; 
it will not mix with the $Z_0$, but it could have a residual  and 
interesting effect to the $\tau$ forward-backward 
asymmetry (it will also give a very 
small shift to the bottom  forward-backward 
asymmetry towards the global average).
At present, 
all the four LEP experiment give an excess in the $A^\tau_\mt{FB}$, 
but the average  is still at the level of only $2\sigma$. Since  
 the technique to extract this data is different from the
 $A^\mu_\mt{FB}$ and  $A^e_\mt{FB}$, one could also suspect that 
the systematic error in this measurement is underestimated.
Before concluding, we also point out that the existence of new generations of 
gauge bosons
 implies additional massive gluons which could have an interesting role 
in explaining the CDF excess in the jet high $p_t$ distribution [5].
\begin{ack}
It is a pleasure to thank G.Altarelli, R.Barbieri and G.G.Ross for
very interesting discussions and helpful comments.
\end{ack}
 
\end{document}